\documentclass[fleqn,12pt,twoside]{article}
\usepackage[headings]{espcrc1}

\readRCS
$Id: espcrc1.tex,v 1.2 2004/02/24 11:22:11 spepping Exp $
\usepackage{graphicx}
\usepackage[figuresright]{rotating}

\newcommand{\AmS}{{\protect\the\textfont2
  A\kern-.1667em\lower.5ex\hbox{M}\kern-.125emS}}

\title{\centering A Global Model of $\beta^-$-Decay Half-Lives Using Neural Networks}

\author{\centering N.~Costiris,\address[UOA]{\small Physics Department, Division of Nuclear Physics and Particle Physics,\\
University of Athens, GR-15771 Athens, Greece}      
        E.~Mavrommatis,\addressmark[UOA]      
        K.~A.~Gernoth,\address{\small Department of Physics, UMIST, P.O. Box 88, Manchester M60 1QD,\\
United Kingdom
} 
        and 
        J.~W.~Clark\address{\small McDonnell Center for the Space Sciences and Department of Physics,\\
Washington University, St. Louis, Missouri 63130, USA}}
       
\runtitle{\small A Global Model of $\beta^-$ Decay Half-Lives Using Neural Networks}
\runauthor{\small N. Costiris}

\begin{document}

\maketitle

\begin{abstract}
\small
Statistical modeling of nuclear data using artificial neural networks 
(ANNs) and, more recently, support vector machines (SVMs), is providing 
novel approaches to systematics that are complementary to phenomenological
and semi-microscopic theories.  We present a global model of 
$\beta^-$-decay halflives of the class of nuclei that decay 100\% 
by $\beta^-$ mode in their ground states. A fully-connected multilayered 
feed forward network has been trained using the Levenberg-Marquardt 
algorithm, Bayesian regularization, and cross-validation.  The halflife
estimates generated by the model are discussed and compared with the
available experimental data, with previous results obtained with 
neural networks, and with estimates coming from traditional global 
nuclear models.  Predictions of the new neural-network model are
given for nuclei far from stability, with particular attention
to those involved in r-process nucleosynthesis.  This study 
demonstrates that in the framework of the $\beta^-$-decay problem
considered here, global models based on ANNs can at least match 
the predictive performance of the best conventional global models 
rooted in nuclear theory.  Accordingly, such statistical models 
can provide a valuable tool for further mapping of the nuclidic 
chart.
\end{abstract}

\section{Introduction}

Currently, there is an urgent need for reliable estimates of 
$\beta^-$-decay halflives of nuclei far from stability.  This
need is driven both by the experimental programs of existing 
and future radioactive ion-beam facilities and by ongoing
efforts toward understanding supernova explosions and the 
processes of nucleosynthesis in stars, notably the r-process [1].  
Such estimates are also needed to provide guidance for nuclear 
structure theory itself, as totally new areas of the nuclear 
landscape are opened for exploration.  Several models for 
determining $\beta^-$ halflives  have been proposed and applied 
during the last few decades. These include the more phenomenological
models based on Gross Theory as well as  models (in various versions)
that employ the Quasiparticle Random-Phase Approximation (QRPA), 
along with some approaches based on shell-model calculations.
The latest version of Gross Theory, known as the Semi-Gross 
Theory (SGT), incorporates shell effects of the parent nucleus
[2].  Extensive proton-neutron ($pn$) QRPA calculations
have been carried out by two groups.  The latest version
of the models developed by the first group (Klapdor and
coworkers) takes account of $pp$ and $pn$ forces and 
includes a schematic interaction for the first-forbidden (\textit{ff})
$\beta$ decay (NBSC+\textit{pn}QRPA) [3]. The latest version of the models developed
by the second group (M\"oller and coworkers) combines the
$pn$QRPA model with the statistical Gross Theory of \textit{ff}
decay (\textit{pn}QRPA+\textit{ff}GT) [4].  There is also a model of $\beta^-$-decay halflives 
in which the ground state of the parent nucleus is described by 
the extended Thomas-Fermi plus Strutinsky integral method and the
continuum QRPA (CQRPA) [5].  Recently, a relativistic $pn$QRPA 
model has been applied in the treatment of neutron-rich nuclei 
in the $N = 50$ and 82 regions [6].  Although there is continuous 
improvement, the predictive power of these conventional models 
is rather limited for $\beta^-$-decay halflives of nuclei 
that are mainly far from the stability, with deviations
from experiment of at least an order of magnitude.  This being
the case, statistical modeling based on artificial neural 
networks (ANNs) and other adaptive techniques of statistical 
inference presents an interesting and potentially effective
alternative for global modeling of $\beta$-decay lifetimes,
as it does for other problems in nuclear systematics.  
Neural-network models have already been developed for
other nuclear properties, including atomic masses,
neutron separation energies, ground-state spins and 
parities, and branching probabilities in different 
decay channels, as well as $\beta^-$ halflives [7]. Very 
recently, global statistical models of some of these
properties have also been developed based on Support Vector 
Machines (SVMs) [8].  In the present work [9], which continues
a previous effort in statistical modeling of nuclear
half-life systematics [7,10,11], we present a new global model for the
halflives of nuclear ground states that decay 
100\% by the $\beta^-$ mode.  The model is a feedforward
artificial neural network (FF-ANN) which has been constructed by
means of a more sophisticated technology than applied previously.
The predictive power of this model is as good or better than
that of the existing models created within nuclear theory.
The methodology adopted is outlined in Sect.~2.  Some results
from the model are reported and evaluated in Sect.~3, which 
is followed in Sect.~4 by a brief summary and prospectus.

\section{Methodology}

After a large number of computer experiments on networks constructed 
with various choices of architectures, training and initialization 
algorithms, forms of activation function, and scaling of variables, 
we have arrived at an ANN model that is effective not only in 
approximating the observed $\beta^-$ decay half-life systematics, 
but also generalizes quite well to unknown regions of the nuclidic 
chart.  This model is a fully connected feedforward (FF) multilayer 
network model with architecture symbolized by $[3-5-5-5-5-1|116]$.  
The activation functions of the neuron-like units are given
by a hyperbolic-tangent sigmoid function in the intermediate 
(hidden) layers and a saturated linear function in the output layer. 
The three input units encode the atomic number $Z$, the neutron 
number $N$, and the corresponding pairing constant 
$\delta$ defined as
\begin{eqnarray}
\delta  = \left\{ {\begin{array}{*{20}c}
   { + 1,}  \\
   {0,}  \\
   { - 1,}  \\
\end{array}\begin{array}{*{20}c}
   {\begin{array}{*{20}c}
   {\rm for} & {\rm even-even} & {\rm nuclei}\,,  \\
\end{array}}  \\
   {\begin{array}{*{20}c}
   {\rm for} & {\rm odd-mass} & {\rm nuclei}\,,  \\
\end{array}}  \\
   {\begin{array}{*{20}c}
   {\rm for} & {\rm odd-odd} & {\rm nuclei}\,.  \\
\end{array}}  \\
\end{array}} \right.
\label{eq:228}
\end{eqnarray}
When $Z$, $N$, and $\delta$ are fed into the input interface, the 
network performs its calculation and the single output unit decodes the 
network's response -- i.e., its estimate for $\log T_{1/2}$,
the base-10 log of the $\beta^-$ halflife of nuclide $(Z,N)$.
The four hidden layers, each containing five neuronal units, 
transfer information from input to output, processing it through
weighted connections (or biases), in this case 116 in number.  
The central idea of neural-network modeling is that such parameters 
can be adjusted through a proper training algorithm to produce good 
overall performance on a training (or learning) set and good 
generalization on test nuclei absent from the training set.  
We have adopted the Levenberg-Marquardt 
optimization algorithm to train the network, while implementing a 
combination of two well-established techniques for improving 
generalization, namely, {\it Bayesian regularization} and 
{\it cross-validation} [12].  The Nguyen-Widrow method~\cite{N1} was 
chosen for initialization of the network.
The experimental data for our modeling of $\beta^-$ halflives have 
been taken from the Nubase2003 evaluation of nuclear and decay 
properties due to G. Audi et al.~\cite{A6}.  Considering only those
cases in which the parent ground state decays 100\% by the $\beta^-$ 
channel, we form a preliminary set made up of 905 nuclides.  The halflives 
of nuclides in this set range from $0.15 \times 10^{-2}\,{\rm s}$ for 
$^{35}$Na to $2.43\times 10^{23}\,{\rm s}$ for $^{113}$Cd.  We have worked 
mostly with a restricted set of 843 nuclides, formed by elimination 
of those nuclei having halflife greater than $10^6\,{\rm s}$.  With the 
exclusion of these long-lived examples, one is dealing with a smaller 
but more homogeneous collection of nuclides that facilitates the 
training of the network.  From the 843 nuclides (\textit{overall set}) (sorted according to the value of the half-life), 503 ($\sim$ 60\%) have been \textit{uniformly} selected to
serve as training examples (\textit{learning set}); of those left
over, 167 ($\sim$ 20\%) have been also \textit{uniformly} chosen to validate the learning procedure (\textit{validation set});
and the remaining 168 nuclides ($\sim$ 20\%) are reserved to test
the accuracy of prediction (\textit{prediction set}).  This 
partitioning of the data was adopted to ensure that the distribution
over halflives in the whole set is faithfully reflected in the learning, validation, and prediction sets.

The performance of the statistical model so designed and constructed
is evaluated with the aid of two commonly used statistical metrics, 
namely the root-mean-square error (RMSE) and the mean-absolute-square
error (MAE):
\begin{equation}
{\rm RMSE} = \left[ {\frac{1}{N_\alpha}\sum\limits_k 
{\left( {y_k  - \hat y_k } \right)^2 } } \right]^{1/2}\,, \qquad 
{\rm MAE} = \frac{1}{N_\alpha}\sum\limits_k {\left| 
{y_k  - \hat y_k } \right|}\,.
\label{eq:29}
\end{equation}
Here, ${\hat y_k}$ is the experimentally measured value of the
base-10 logarithm of the halflife of nuclide $k$,
and $y_k$ is the estimate of this quantity (the ``calculated
value'') produced by the network model.  The sums in definitions
(\ref{eq:29}) run over the $N_\alpha$ nuclides in the learning, validation,
or prediction set, as appropriate.  Smaller values of these metrics
indicate higher accuracy.  
In the training procedure adopted, the RMSE was taken as the
cost function, or objective function, to be minimized by adjustment 
of the weight parameters of the network.

In the literature on global modeling of $\beta^-$ lifetimes, several
different figures of merit have been used to analyze model performance.
The collaboration led by Klapdor \cite{M1} employs the average
deviation defined by
\begin{equation}
\langle x \rangle_K  = \frac{1}{N_\alpha}\sum\limits_i {x_i }\,,
\label{eq:39}
\end{equation}
where $x_i$ is given by 
$T_{{1/2},{({\rm exp})}}^{(i)}/T_{{1/2},{({\rm calc})}}^{(i)}$
if $T_{{1/2},{({\rm exp})}}^{(i)} \geq T_{{1/2}^{({\rm calc})}}^{(i)}$, 
the ratio otherwise being inverted, along with the corresponding 
standard deviation 
\begin{equation}
\sigma _K  = \left[ {\frac{1}{N_\alpha}\sum\limits_i {\left( {x_i  
- \langle x \rangle_K } \right)^2 } } \right]^{{1 \mathord{\left/
 {\vphantom {1 2}} \right.
 \kern-\nulldelimiterspace} 2}} .
\label{eq:43}
\end{equation}
(Again, the sums run over the appropriate set of nuclides.)
Following Klapdor and coworkers,
a more incisive analysis is achieved by determining the percentage 
$m$ of nuclides having measured ground-state halflife 
$T_{{1 \mathord{\left/ {\vphantom {1 {2,{\rm exp} }}} \right. 
\kern-\nulldelimiterspace} {2,{\rm exp} }}}$ within a prescribed range 
(e.g., not greater than $10^6$ $s$, 60 $s$, or 1 $s$), for which the 
calculated halflife is within a prescribed tolerance factor $f$ 
(in particular 2, 5, or 10) of the experimental value.  
M\"{o}ller et al.~\cite{M3} have used a measure $M_r$ similar 
to $\langle x \rangle_K$, but defined in terms of $\log_{10}T_{1/2}$
rather than $T_{1/2}$.  Thus, they analyzed model performance in
terms of the mean value of $r_k = y_k/{\hat y_k}$ and its associated
standard deviation, given respectively by
\begin{equation}
M_r = \frac{1}{N_\alpha} \sum_k r_k \quad {\rm and} \quad  
\sigma_r = \left[ \frac{1}{N_\alpha} \sum_k ( r_k - M_r )^2 \right]^{1/2} \,.
\label{eq:45}
\end{equation}
Somewhat more useful are the mean deviation (range) and mean fluctuation
(range), defined respectively as 
\begin{equation}
M^{(10)} = 10^{M_r} \quad {\rm and} \quad \sigma_{M^{(0)}} =
10^{\sigma_r} \,.
\label{eq:49}
\end{equation}
For a closer analysis, these indices may again be calculated within
prescribed halflife ranges.

\section{Results}

Table~\ref{table:0} presents the RMSE and the MAE of our global 
model on the indicated data sets.  Also included for comparison
are some results from an antecedent ANN model~\cite{A7}.  The 
improvement represented by the current model is apparent.
  
\begin{table}[htb]
\caption{\small Root-mean-square errors (RMSE) and mean-absolute errors 
(MAE) (Eqs.~(\ref{eq:29})) for learning, validation, and prediction 
subsets and the full data set (overall set), for current and previous
FF-ANN models of $\beta^{-}$ lifetime systematics [10].}
\label{table:0}
\newcommand{\m}{\hphantom{$-$}}
\newcommand{\cc}[1]{\multicolumn{1}{c}{#1}}
\renewcommand{\tabcolsep}{1.8pc} 
\begin{tabular}{lllll}
\hline
\multicolumn{3}{c}{\small \bf Current FF-ANN Model}  &\multicolumn{2}{c}{\small \bf Previous FF-ANN Model~\cite{A7}} \\ 
\multicolumn{3}{c}{\small $[3-5-5-5-5-1|116]$}  &\multicolumn{2}{c}{\small $[16-10-1|181]$ } \\ 
\hline
\multicolumn{1}{c}{\small \bf Data Set} & \multicolumn{1}{c}{\small RMSE} &  \multicolumn{1}{c}{\small MAE}  &\multicolumn{1}{c}{\small \bf Data Set} & \multicolumn{1}{c}{\small RMSE} \\ 
\hline
\multicolumn{1}{c}{\small \bf Learning} & \multicolumn{1}{c}{\small 0.53} &  \multicolumn{1}{c}{\small 0.38}  &\multicolumn{1}{c}{\small \bf Learning} & \multicolumn{1}{c}{\small 1.08} \\ 
\multicolumn{1}{c}{\small \bf Validation} & \multicolumn{1}{c}{\small 0.60} &   \multicolumn{1}{c}{\small 0.41}  &\multicolumn{1}{c}{\small -} & \multicolumn{1}{c}{\small -} \\ 
\multicolumn{1}{c}{\small \bf Prediction} & \multicolumn{1}{c}{\small 0.65} &   \multicolumn{1}{c}{\small 0.46}  &\multicolumn{1}{c}{\small \bf Prediction} & \multicolumn{1}{c}{\small 1.82} \\ 
\multicolumn{1}{c}{\small \bf Overall} & \multicolumn{1}{c}{\small \bf 0.57} &   \multicolumn{1}{c}{\small \bf 0.40}  &\multicolumn{1}{c}{\small -} & \multicolumn{1}{c}{\small -} \\ 
\hline
\end{tabular}
\\[2pt]
\end{table}

Next, adopting the performance metric of M\"oller and collaborators
(Eqs.~(\ref{eq:45})-(\ref{eq:49})), our model is compared with two
global models based on the proton-neutron Quasiparticle
Random-Phase Approximation ($pn$QRPA), namely the NBSC+$pn$QRPA model of 
Homma et~al.~\cite{M1} and the FRDM+$pn$QRPA model of M\"{o}ller 
et~al.~\cite{M2}.  Table~\ref{table:2} contains the results for $M^{(10)}$ 
and $\sigma_{M^{(10)}}$, according to the even-even, odd-odd, and 
odd-mass-number character of nuclides, excluding halflives greater than 
1000 s.  As seen in the table, the $pn$QRPA models tend to overestimate 
halflives for odd-odd nuclei (except for the shorter-lived nuclides
in the FRDM+$pn$QRPA case). The FRDM+$pn$QRPA model also overestimates 
halflives for even-even nuclei.  Our model also tends to overestimate 
the halflives of even-even nuclei, but to a substantially smaller
degree.  This shortcoming of the ANN model is partly due to 
the more limited abundance of these nuclei.

\begin{table}[htb]
\caption{\small Values of the performance measures $M^{(10)}$ and 
$\sigma_{M^{(10)}}$ (Eqs.~(\ref{eq:45})-(\ref{eq:49})) for the current FF-ANN model (overall and prediction sets) and for the NBSC+\textit{pn}QRPA~\cite{M1} and FRDM+\textit{pn}QRPA~\cite{M2} models.}
\label{table:2}
\newcommand{\m}{\hphantom{$-$}}
\newcommand{\cc}[1]{\multicolumn{1}{c}{#1}}
\renewcommand{\tabcolsep}{1.2pc} 
\begin{tabular}{llllllll}
\hline
\multicolumn{1}{c}{\small $\bf T_{{1 \mathord{\left/ {\vphantom {1 {2,{\bf exp} }}} \right. \kern-\nulldelimiterspace} {2,{\rm exp} }}}$} &  & \multicolumn{3}{c}{\small \bf Overall Set} & \multicolumn{3}{c}{\small \bf Prediction Set} \\ 
\multicolumn{1}{c}{\small (sec)} & \multicolumn{1}{c}{\small char.} & \multicolumn{1}{c}{\small $n$} & \multicolumn{1}{c}{\small $M^{(10)}$} & \multicolumn{1}{c}{\small $\sigma_{M^{(10)}}$} & \multicolumn{1}{c}{\small $n$} & \multicolumn{1}{c}{\small $M^{(10)}$} & \multicolumn{1}{c}{\small $\sigma_{M^{(10)}}$} \\ 
\hline
\small $f <1$ & \multicolumn{1}{c}{\small o-o} & \multicolumn{1}{c}{\small 76} & \multicolumn{1}{c}{\small 1.04} & \multicolumn{1}{c}{\small 2.53} & \multicolumn{1}{c}{\small 11} & \multicolumn{1}{c}{\small 0.86} & \multicolumn{1}{c}{\small 1.98} \\ 
 & \multicolumn{1}{c}{\small odd} & \multicolumn{1}{c}{\small 125} & \multicolumn{1}{c}{\small 1.16} & \multicolumn{1}{c}{\small 2.25} & \multicolumn{1}{c}{\small 32} & \multicolumn{1}{c}{\small 1.05} & \multicolumn{1}{c}{\small 2.40} \\ 
 & \multicolumn{1}{c}{\small e-e} & \multicolumn{1}{c}{\small 51} & \multicolumn{1}{c}{\small 1.87} & \multicolumn{1}{c}{\small 2.45} & \multicolumn{1}{c}{\small 7} & \multicolumn{1}{c}{2.36} & \multicolumn{1}{c}{\small 3.26} \\ 
\small $f <10$ & \multicolumn{1}{c}{\small o-o} & \multicolumn{1}{c}{\small 121} & \multicolumn{1}{c}{\small 1.11} & \multicolumn{1}{c}{\small 2.96} & \multicolumn{1}{c}{\small 20} & \multicolumn{1}{c}{\small 0.86} & \multicolumn{1}{c}{\small 3.76} \\ 
 & \multicolumn{1}{c}{\small odd} & \multicolumn{1}{c}{\small 187} & \multicolumn{1}{c}{\small 1.10} & \multicolumn{1}{c}{\small 2.31} & \multicolumn{1}{c}{\small 42} & \multicolumn{1}{c}{\small 0.92} & \multicolumn{1}{c}{\small 2.61} \\ 
 & \multicolumn{1}{c}{\small e-e} & \multicolumn{1}{c}{\small 87} & \multicolumn{1}{c}{\small 1.65} & \multicolumn{1}{c}{\small 2.56} & \multicolumn{1}{c}{\small 17} & \multicolumn{1}{c}{\small 1.80} & \multicolumn{1}{c}{\small 2.58} \\ 
\small $f < 100$ & \multicolumn{1}{c}{\small o-o} & \multicolumn{1}{c}{\small 158} & \multicolumn{1}{c}{\small 1.08} & \multicolumn{1}{c}{\small 3.06} & \multicolumn{1}{c}{\small 28} & \multicolumn{1}{c}{\small 0.76} & \multicolumn{1}{c}{\small 3.20} \\ 
 & \multicolumn{1}{c}{\small odd} & \multicolumn{1}{c}{\small 261} & \multicolumn{1}{c}{\small 1.08} & \multicolumn{1}{c}{\small 2.45} & \multicolumn{1}{c}{\small 57} & \multicolumn{1}{c}{0.97} & \multicolumn{1}{c}{2.91} \\ 
 & \multicolumn{1}{c}{\small e-e} & \multicolumn{1}{c}{\small 110} & \multicolumn{1}{c}{\small 1.58} & \multicolumn{1}{c}{\small 2.31} & \multicolumn{1}{c}{\small 21} & \multicolumn{1}{c}{\small 1.58} & \multicolumn{1}{c}{\small 2.98} \\ 
\small $f<1000$ & \multicolumn{1}{c}{\small o-o} & \multicolumn{1}{c}{\small 191} & \multicolumn{1}{c}{\small 1.12} & \multicolumn{1}{c}{\small 3.06} & \multicolumn{1}{c}{\small 35} & \multicolumn{1}{c}{0.78} & \multicolumn{1}{c}{\small 3.13} \\ 
 & \multicolumn{1}{c}{\small odd} & \multicolumn{1}{c}{\small 329} & \multicolumn{1}{c}{\small 1.07} & \multicolumn{1}{c}{\small 2.73} & \multicolumn{1}{c}{\small 68} & \multicolumn{1}{c}{\small 0.84} & \multicolumn{1}{c}{\small 3.07} \\ 
 & \multicolumn{1}{c}{\small e-e} & \multicolumn{1}{c}{\small 133} & \multicolumn{1}{c}{\small 1.63} & \multicolumn{1}{c}{\small 2.60} & \multicolumn{1}{c}{\small 28} & \multicolumn{1}{c}{\small 1.49} & \multicolumn{1}{c}{\small 3.04} \\

\hline
\multicolumn{1}{c}{\small $\bf T_{{1 \mathord{\left/ {\vphantom {1 {2,{\bf exp} }}} \right. \kern-\nulldelimiterspace} {2,{\rm exp} }}}$} &  & \multicolumn{3}{c}{\small \bf NBSC+\textit{pn}QRPA~\cite{M1}} & \multicolumn{3}{c}{\small \bf FRDM+\textit{pn}QRPA~\cite{M2}} \\ 
\multicolumn{1}{c}{\small (sec)} & \multicolumn{1}{c}{\small char.} & \multicolumn{1}{c}{\small $n$} & \multicolumn{1}{c}{\small $M^{(10)}$} & \multicolumn{1}{c}{\small $\sigma_{M^{(10)}}$} & \multicolumn{1}{c}{\small $n$} & \multicolumn{1}{c}{\small $M^{(10)}$} & \multicolumn{1}{c}{\small $\sigma_{M^{(10)}}$} \\ 
\hline
\small $f <1$ & \multicolumn{1}{c}{\small o-o} & \multicolumn{1}{c}{\small 28} & \multicolumn{1}{c}{\small 1.75} & \multicolumn{1}{c}{\small 4.96} & \multicolumn{1}{c}{\small 29} & \multicolumn{1}{c}{\small 0.59} & \multicolumn{1}{c}{\small 2.91} \\ 
 & \multicolumn{1}{c}{\small odd} & \multicolumn{1}{c}{\small 31} & \multicolumn{1}{c}{\small 0.60} & \multicolumn{1}{c}{\small 2.24} & \multicolumn{1}{c}{\small 35} & \multicolumn{1}{c}{\small 0.59} & \multicolumn{1}{c}{\small 2.64} \\ 
 & \multicolumn{1}{c}{\small e-e} & \multicolumn{1}{c}{\small 10} & \multicolumn{1}{c}{\small 1.15} & \multicolumn{1}{c}{\small 2.36} & \multicolumn{1}{c}{\small 10} & \multicolumn{1}{c}{\small 3.84} & \multicolumn{1}{c}{\small 3.08} \\ 
\small $f <10$ & \multicolumn{1}{c}{\small o-o} & \multicolumn{1}{c}{\small 66} & \multicolumn{1}{c}{\small 1.89} & \multicolumn{1}{c}{\small 4.60} & \multicolumn{1}{c}{\small 59} & \multicolumn{1}{c}{0.76} & \multicolumn{1}{c}{\small 8.83} \\ 
 & \multicolumn{1}{c}{\small odd} & \multicolumn{1}{c}{\small 81} & \multicolumn{1}{c}{\small 0.92} & \multicolumn{1}{c}{\small 3.84} & \multicolumn{1}{c}{\small 85} & \multicolumn{1}{c}{0.78} & \multicolumn{1}{c}{\small 4.81} \\ 
 & \multicolumn{1}{c}{\small e-e} & \multicolumn{1}{c}{\small 34} & \multicolumn{1}{c}{\small 1.01} & \multicolumn{1}{c}{\small 2.93} & \multicolumn{1}{c}{\small 34} & \multicolumn{1}{c}{2.50} & \multicolumn{1}{c}{\small 4.13} \\ 
\small $f <100$ & \multicolumn{1}{c}{\small o-o} & \multicolumn{1}{c}{\small 85} & \multicolumn{1}{c}{\small 3.15} & \multicolumn{1}{c}{\small 10.5} & \multicolumn{1}{c}{\small 88} & \multicolumn{1}{c}{\small 2.33} & \multicolumn{1}{c}{\small 49.2} \\ 
 & \multicolumn{1}{c}{\small odd} & \multicolumn{1}{c}{\small 127} & \multicolumn{1}{c}{\small 1.07} & \multicolumn{1}{c}{\small 4.29} & \multicolumn{1}{c}{\small 133} & \multicolumn{1}{c}{\small 1.11} & \multicolumn{1}{c}{\small 9.45} \\ 
 & \multicolumn{1}{c}{\small e-e} & \multicolumn{1}{c}{\small 52} & \multicolumn{1}{c}{\small 1.13} & \multicolumn{1}{c}{\small 3.58} & \multicolumn{1}{c}{\small 54} & \multicolumn{1}{c}{\small 2.61} & \multicolumn{1}{c}{\small 4.75} \\ 
\small $f <1000$ & \multicolumn{1}{c}{\small o-o} & \multicolumn{1}{c}{\small 93} & \multicolumn{1}{c}{\small 3.02} & \multicolumn{1}{c}{\small 10.2} & \multicolumn{1}{c}{\small 115} & \multicolumn{1}{c}{\small 3.50} & \multicolumn{1}{c}{\small 72.0} \\ 
 & \multicolumn{1}{c}{\small odd} & \multicolumn{1}{c}{\small 157} & \multicolumn{1}{c}{\small 1.10} & \multicolumn{1}{c}{\small 5.55} & \multicolumn{1}{c}{\small 194} & \multicolumn{1}{c}{\small 2.77} & \multicolumn{1}{c}{\small 71.5} \\ 
 & \multicolumn{1}{c}{\small e-e} & \multicolumn{1}{c}{\small 63} & \multicolumn{1}{c}{\small 1.39} & \multicolumn{1}{c}{\small 6.10} & \multicolumn{1}{c}{\small 71} & \multicolumn{1}{c}{\small 6.86} & \multicolumn{1}{c}{\small 58.5} \\ 
\hline
\end{tabular}\\[2pt]
\end{table}

The efficacy of our latest FF-ANN model has also been evaluated
in terms of the metrics introduced by Klapdor and coworkers
(Eqs.~(\ref{eq:39})-(\ref{eq:43})).  Table~\ref{table:1} includes 
our results as well as those of the NBSC+$pn$QRPA model~\cite{M1}.  
Comparing the two models, especially the values for the percentage
$m$, it is evident that the neural-network model is superior
in estimating the $\beta^-$ lifetimes, both for the totality
of the data and for the prediction set.  The ANN model has the
ability to reproduce approximately 90\% of experimentally known halflives 
shorter than $10^6\,{\rm s}$ within a factor of 10 and about 
50\% within a factor of 2.  That $m$ is reduced only slightly 
in going from the overall set to the prediction set is indicative 
of good generalization.  However, the way in which the three 
subsets (for training, validation, and testing) were chosen 
dictates that the prediction involved in generating the
results of Table~\ref{table:1} is primarily a matter of
interpolation rather than extrapolation.

\begin{table}[htb]
\caption{\small Values of the performances measures $\langle x \rangle_K$
and $\sigma_K$ (Eqs.~(\ref{eq:39})-(\ref{eq:43})) for our FF-ANN model 
(overall and prediction sets) and for the NBSC+$pn$QRPA model~\cite{M1}.
Also given are results for the percentage $m$ of nuclides having
measured halflife within the prescribed range, for which the halflife
given by the model lies within a tolerance factor $f$ of the
experimental value.}
\label{table:1}
\newcommand{\m}{\hphantom{$-$}}
\newcommand{\cc}[1]{\multicolumn{1}{c}{#1}}
\begin{tabular}{lllllllllll}

\hline
\multicolumn{1}{c}{} & \multicolumn{1}{c}{\small $\bf T_{{1 \mathord{\left/ {\vphantom {1 {2,{\bf exp}}}} \right. \kern-\nulldelimiterspace} {2,{\rm exp} }}}$} & \multicolumn{3}{c}{\small \bf Overall Set} & \multicolumn{3}{c}{\small \bf Prediction Set} & \multicolumn{3}{c}{\small \bf NBSC+\textit{pn}QRPA~\cite{M1}} \\ 
\multicolumn{1}{c}{\small $f$} & \multicolumn{1}{c}{\small (sec)} & \multicolumn{1}{c}{\small m(\%)} & \multicolumn{1}{c}{\small $\langle x\rangle_K$} & \multicolumn{1}{c}{\small $\sigma_K$} & \multicolumn{1}{c}{\small m(\%)} & \multicolumn{1}{c}{\small $\langle x\rangle_K$} & \multicolumn{1}{c}{\small $\sigma_K$} & \multicolumn{1}{c}{\small m(\%)} & \multicolumn{1}{c}{\small $\langle x\rangle_K$} & \multicolumn{1}{c}{\small $\sigma_K$} \\ 
\hline
\small $f <10 $ & \small $<10^6$ & \multicolumn{1}{c}{\small 92.0} & \multicolumn{1}{c}{\small 2.46} & \multicolumn{1}{c}{\small 1.72} & \multicolumn{1}{c}{\small 90.5} & \multicolumn{1}{c}{\small 2.69} & \multicolumn{1}{c}{\small 1.85} & \multicolumn{1}{c}{\small 76.7} & \multicolumn{1}{c}{\small 3.00} & \multicolumn{1}{c}{\small -} \\ 
á  & \small $<60$ & \multicolumn{1}{c}{\small 96.5} & \multicolumn{1}{c}{\small 2.21} & \multicolumn{1}{c}{\small 1.52} & \multicolumn{1}{c}{\small 96.1} & \multicolumn{1}{c}{\small 2.48} & \multicolumn{1}{c}{\small 1.64} & \multicolumn{1}{c}{\small 87.2} & \multicolumn{1}{c}{\small 2.81} & \multicolumn{1}{c}{\small -} \\ 
 & \small $f <1$ & \multicolumn{1}{c}{\small 97.6} & \multicolumn{1}{c}{\small 2.10} & \multicolumn{1}{c}{\small 1.39} & \multicolumn{1}{c}{\small 98.0} & \multicolumn{1}{c}{\small 2.24} & \multicolumn{1}{c}{\small 1.30} & \multicolumn{1}{c}{\small 95.7} & \multicolumn{1}{c}{\small 2.64} & \multicolumn{1}{c}{\small -} \\ 
\small $f <5 $ & \small $<10^6$ & \multicolumn{1}{c}{\small 82.8} & \multicolumn{1}{c}{\small 1.99} & \multicolumn{1}{c}{\small 0.95} & \multicolumn{1}{c}{\small 79.2} & \multicolumn{1}{c}{\small 2.10} & \multicolumn{1}{c}{\small 0.97} & \multicolumn{1}{c}{\small -} & \multicolumn{1}{c}{\small -} & \multicolumn{1}{c}{\small -} \\ 
á  & \small $<60$ & \multicolumn{1}{c}{\small 90.2} & \multicolumn{1}{c}{\small 1.88} & \multicolumn{1}{c}{\small 0.84} & \multicolumn{1}{c}{\small 87.3} & \multicolumn{1}{c}{\small 2.05} & \multicolumn{1}{c}{\small 0.91} & \multicolumn{1}{c}{\small -} & \multicolumn{1}{c}{\small -} & \multicolumn{1}{c}{\small -} \\ 
 & \small $f <1$ & \multicolumn{1}{c}{\small 93.7} & \multicolumn{1}{c}{\small 1.88} & \multicolumn{1}{c}{\small 0.80} & \multicolumn{1}{c}{\small 94.0} & \multicolumn{1}{c}{\small 2.04} & \multicolumn{1}{c}{\small 0.89} & \multicolumn{1}{c}{\small -} & \multicolumn{1}{c}{\small -} & \multicolumn{1}{c}{\small -} \\ 
\small $f <2 $ & \small $<10^6$ & \multicolumn{1}{c}{\small 53.5} & \multicolumn{1}{c}{\small 1.41} & \multicolumn{1}{c}{\small 0.27} & \multicolumn{1}{c}{\small 49.4} & \multicolumn{1}{c}{\small 1.48} & \multicolumn{1}{c}{\small 0.28} & \multicolumn{1}{c}{\small 33.8} & \multicolumn{1}{c}{\small 1.43} & \multicolumn{1}{c}{\small -} \\ 
 & \small $<60$ & \multicolumn{1}{c}{\small 60.6} & \multicolumn{1}{c}{\small 1.41} & \multicolumn{1}{c}{\small 0.27} & \multicolumn{1}{c}{\small 53.9} & \multicolumn{1}{c}{\small 1.48} & \multicolumn{1}{c}{\small 0.27} & \multicolumn{1}{c}{\small 42.0} & \multicolumn{1}{c}{\small 1.41} & \multicolumn{1}{c}{\small -} \\ 
 & \small $f<1$ & \multicolumn{1}{c}{\small 61.9} & \multicolumn{1}{c}{\small 1.41} & \multicolumn{1}{c}{\small 0.26} & \multicolumn{1}{c}{\small 60.0} & \multicolumn{1}{c}{\small 1.50} & \multicolumn{1}{c}{\small 0.27} & \multicolumn{1}{c}{\small 50.7} & \multicolumn{1}{c}{\small 1.43} & \multicolumn{1}{c}{\small -} \\ 
\hline
\end{tabular}\\[2pt]

\end{table}

\begin{figure}[h]
\begin{minipage}[t]{80mm}
\includegraphics[width=80mm]{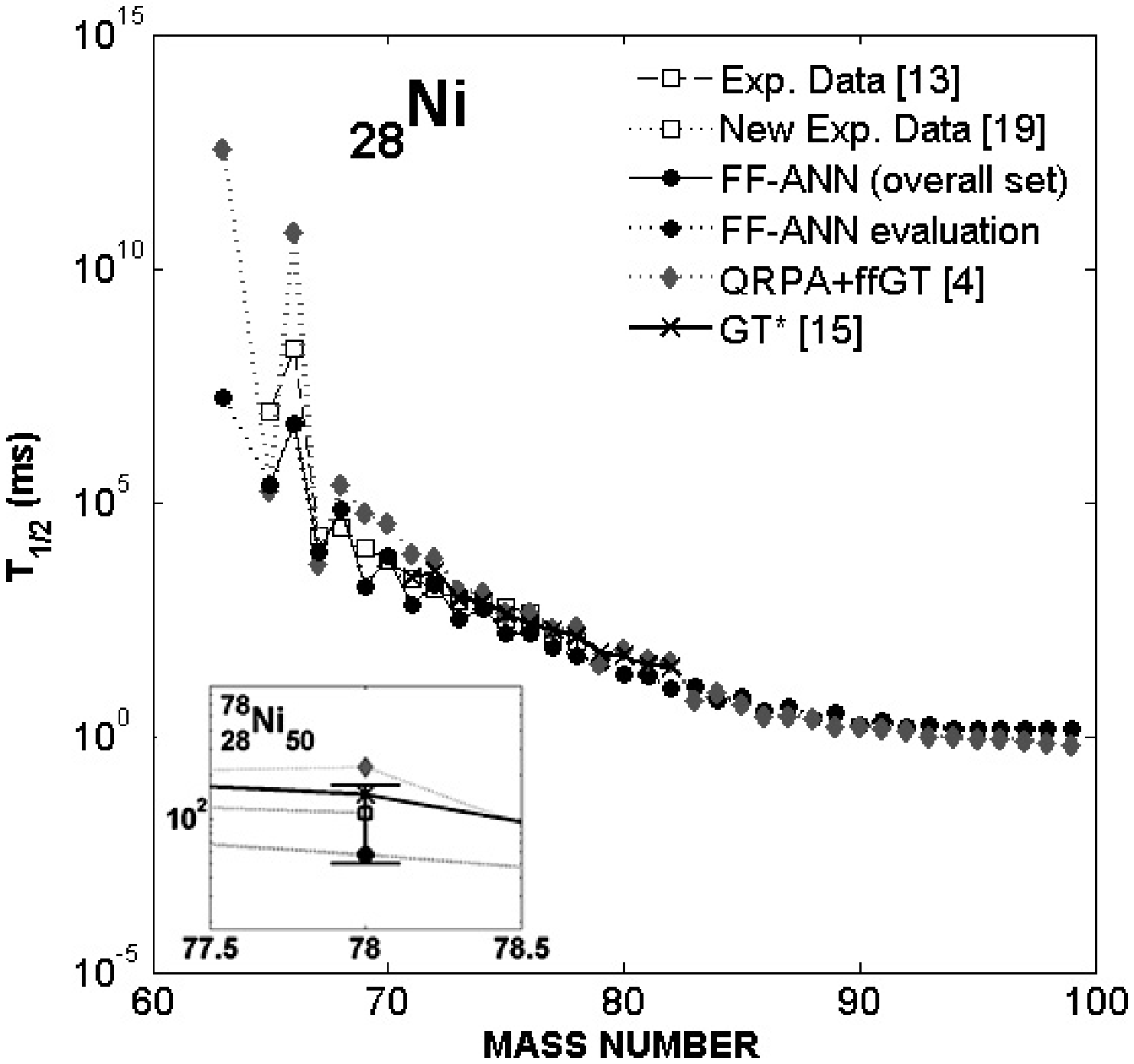}
\caption{\small Halflives $T_{1/2}$ for $\beta^-$ decay of nuclides in 
the isotopic chain of $_{28}$Ni. Included are experimental data 
and results of various models (see legend).  The new measurement 
for the doubly magic r-process nuclide $^{78}_{28}$Ni is included 
(see legend).}
\label{fig:2}
\end{minipage}
\hspace{0.2in}
\begin{minipage}[t]{80mm}
\includegraphics[width=80mm]{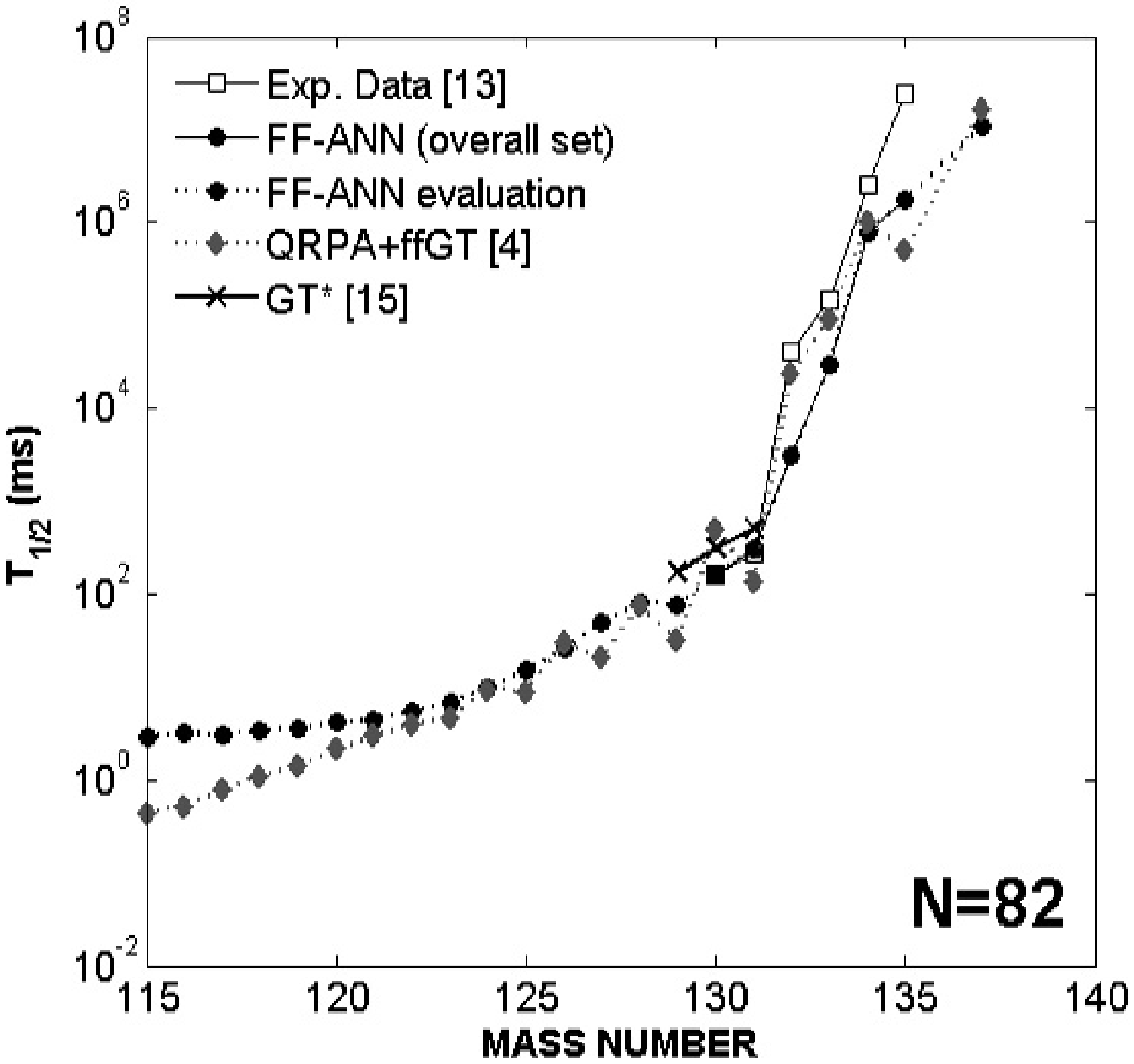}
\caption{\small As in Fig.~\ref{fig:2}, but for the isotonic chain of $N=82$.}
\label{fig:4}
\end{minipage}
\end{figure}

Thus, the capability of the ANN model in interpolation having been
established, we must now assess its potential for extrapolation,
or ``extrapability.''  In this aspect of model behavior, we expect
a similar level of performance as seen in Table~\ref{table:1},
for the early extrapolation regions close in $Z$ and $N$ to 
the learning and validation sets.  
Fig.~\ref{fig:2} shows estimated halflives for 
nuclides in the Ni isotopic chain along with available experimental data, while 
Fig.~\ref{fig:4} presents analogous results for the $N=82$ isotonic 
chain.  For comparison, the plots include corresponding estimates from 
the hybrid $pn$QRPA+\textit{ff}GT model~\cite{M3}, as well as results of some 
calculations by Pfeiffer, Kratz, and M\"{o}ller~\cite{N3} (GT*) based 
on the early Gross Theory (GT) of Takahashi et al.~\cite{M22} 
but with updated mass values~\cite{M24,M23}.  It is interesting to 
compare the various predictions for the halflife of the doubly 
magic r-process $^{78}$Ni nucleus ($N=50$, $Z=28$) with the value 
that was recently measured by Hosmer et al.~\cite{M20} 
(Fig.~\ref{fig:2}). Our result lies within the error bar. 

There is no firm and quantitative guideline by which the
behavior of the different calculations can be judged, either
theoretical or statistical, in the ranges where there are no 
experimental data.  Nevertheless, one does expect, from the 
observed behavior of known nuclei, that the more neutron-rich 
an exotic isotope is, the shorter its halflife will be.  This
projected downward tendency under increasing $N$ is seen in 
all of the models considered.  One does anticipate more drastic
dependence of the halflife on the even/odd character of two
neighboring isotopes.  This sawtooth behavior is present in
our ANN model, but it is probably somewhat exaggerated.  The 
same behavior appears, if to a lesser degree, in 
continuum-Quasiparticle-RPA (CQRPA) approaches~\cite{A23} 
and in other calculations~\cite{M3,M22}.

\section{Conclusion and Prospects}
The results presented here demonstrate that global statistical
models based on Artificial Neural Networks have strong potential
for successful exploitation of the accelerated acquisition of
nuclear data and could provide a valuable tool in exploring
the nature of $\beta^-$ halflives outside the stable valley.
Accordingly, we plan further studies of the systematics of 
$\beta$ decay, employing still more advanced ANN models 
that embody new state-of-the-art optimization algorithms
and more sophisticated training strategies, with the object
of continued enhancement of the predictive power of these
learning machines.

\section{Acknowledgments}
This research has been supported in part by the U. S. National Science Foundation under Grant No. PHY-0140316 and by the University of Athens under Grant No. 70/4/3309.

\end{document}